\documentclass{JAC2003}

\usepackage{cite}
\usepackage{graphicx}
\usepackage{booktabs}

\usepackage{varwidth}

\usepackage{xcolor}

\begin{document}
\title{ LONG-RANGE BEAM--BEAM EFFECTS IN THE LHC}

\author{W. Herr, X. Buffat, R. Calaga, R. Giachino, G. Papotti, T. Pieloni, \\
D. Kaltchev, TRIUMF, Vancouver, Canada}

\maketitle

\begin{abstract}
We report on the experience with long-range beam--beam effects
in the LHC, in dedicated studies as well as the experience from
operation.
Where possible, we compare the observations with the
expectations.
\end{abstract}

\section{LAYOUT FOR BEAM--BEAM INTERACTIONS}
\begin{figure}[htb]
   \centering
   \includegraphics*[width=35mm, height=35mm, angle=-00]{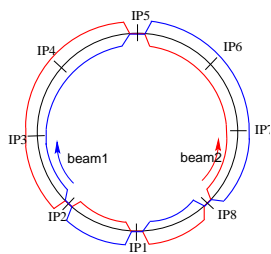}
   \caption{Layout of the experimental collision points in the LHC \cite{wh:01, gp:01}.}
   \label{ref-f0}
\end{figure}

\begin{figure}[htb]
   \centering
   \includegraphics*[width=35mm, height=75mm, angle=-90]{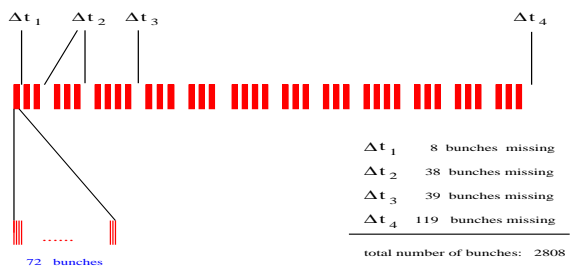}
   \caption{Bunch filling scheme of the nominal LHC.}
   \label{ref-f0a}
\end{figure}

The layout of experimental regions in the LHC is shown in Fig.~\ref{ref-f0}.
The beams travel in separate vacuum chambers and cross in the experimental
areas where they share a common beam pipe.
In these common regions the beams experience head-on collisions as well
as a large number of long-range beam--beam encounters \cite{wh:01, gp:01, wh:02}.
This arrangement together with the bunch filling scheme of the LHC as shown
in Fig.~\ref{ref-f0a} \cite{wh:01, wh:02} leads to very different collision pattern for
different bunches, often referred to as "PACMAN" bunches.
The number of both, head-on as well as long-range encounters, can be
very different for different bunches in the bunch trains and lead to
a different integrated beam--beam effect \cite{gp:01, wh:02}.
This was always a worry in the LHC design and the effects have been
observed in an early stage of the commissioning.
Strategies have been provided
to minimize these effect, e.g.\ different planes for the crossing
angles \cite{wh:01, wh:02}.
\subsection{Strength of Long-range Interactions}
A key parameter for the effect of long-range interactions is the local beam
separation at the parasitic encounters.
Usually this separation is measured in units of the transverse beam size at
the corresponding encounter.
As standard and for comparison between different configurations, we use the {{normalized separation}} in the {\underline{drift space}}. For small enough ${\beta^{*}}$ and round beams it can be written as a simple expression:
\begin{eqnarray}
    d_{sep} \approx \frac{\sqrt{\beta^{*}} \cdot \alpha \cdot \sqrt{\gamma}}{\sqrt{\epsilon_{n}}}~.
\end{eqnarray}
Beyond the drift space the exact separation has to be computed with an optics program.
For a small ${\beta^{*}}$ the phase advance varies fast between the head-on interaction point
and the first parasitic encounter from 0 to $\frac{\pi}{2}$.
Most of the parasitic encounters occur at similar phases and therefore can add up.
This is in strong contrast to other separation schemes such as pretzel separation used in
the Super Proton Synchroton (SPS) collider and the Tevatron, where the parasitic encounters
are distributed around the whole ring.
The separation is easier to control in the case of a crossing angle than for a global
separation.
A strong, local non-linearity can be expected to have a strong effect, but
opens the possibility of a correction.
Possible correction schemes are discussed in a dedicated session at this workshop \cite{comp}.
\section{STUDIES OF LONG-RANGE INTERACTIONS}
Contrary to the head-on beam--beam effects, the long-range beam--beam interactions
are expected to play an important role for the LHC performance and the choice
of the parameters \cite{tp:01, gt:01}.
To study the effect of long-range beam--beam interactions we have performed two dedicated
experiments.
In the first experiment,
the LHC was set up with single trains of 36 bunches per beam, spaced by 50~ns.
The bunch intensities were $\approx~1.2\times 10^{11}$ protons and the normalized
emittances around 2.5~$\mu$m.
The trains collided in IP1 and IP5, leading to a maximum
of 16 long-range encounters per interaction point for nominal bunches.
First, the crossing angle (vertical plane) in IP1 was decreased in small
steps and the losses of each bunch recorded.
The details of this procedure are described in \cite{lr-md}.

In the second experiment we injected 3 trains per beam, with 36 bunches per train.
The filling scheme was chosen such that some trains have collisions in IP1 and IP5
and others collide only in IP2 or IP8.
\subsection{Losses Due to Long-range Interactions}
From simulations \cite{wh:05} we expected a reduction of the dynamic aperture due to the
long-range beam--beam encounters and therefore increased losses when the separation is
decreased.

\begin{figure}[htb]
   \centering
   \includegraphics*[width=35mm, height=75mm, angle=-90]{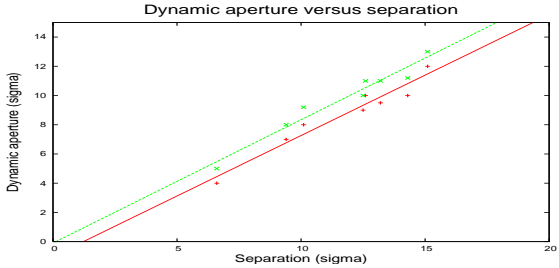}
   \caption{Expected dynamic aperture as function of separation in drift space \cite{wh:05}. Computed for 50~ns and 25~ns separation.}
   \label{ref-f2x}
\end{figure}

To estimate the losses, we show in Fig.~\ref{ref-f2x} the expected dynamic aperture as a function of
the normalized separation \cite{wh:05} for two different bunch spacings (50~ns and 25~ns).
The separation was varied by changing the crossing angle as well as $\beta^{*}~$.
From this figure we can determine that visible (i.e.\ recordable) losses we can expect for a dynamic aperture
around 3~$\sigma$ and therefore when the separation is reduced to values around 5~$\sigma$.
For larger separation the dynamic aperture is also decreased but the losses cannot be observed
in our experiment.

\subsection{Losses Due to Long-range Encounters During Operation}
Significant losses have also been observed during regular operation.
Given that the expected dynamic aperture is closely related to the normalized
separation as shown in Fig.~\ref{ref-f2x}, this separation should be kept
large enough and, if possible, constant during an operation of the machine.
From the expression for the normalized separation it is clear that
a change of $\beta^{*}$ requires a change of the crossing angle $\alpha$ to
keep the separation constant.

\begin{figure}[htb]
   \centering
   \includegraphics*[width=35mm, height=105mm, angle=-90]{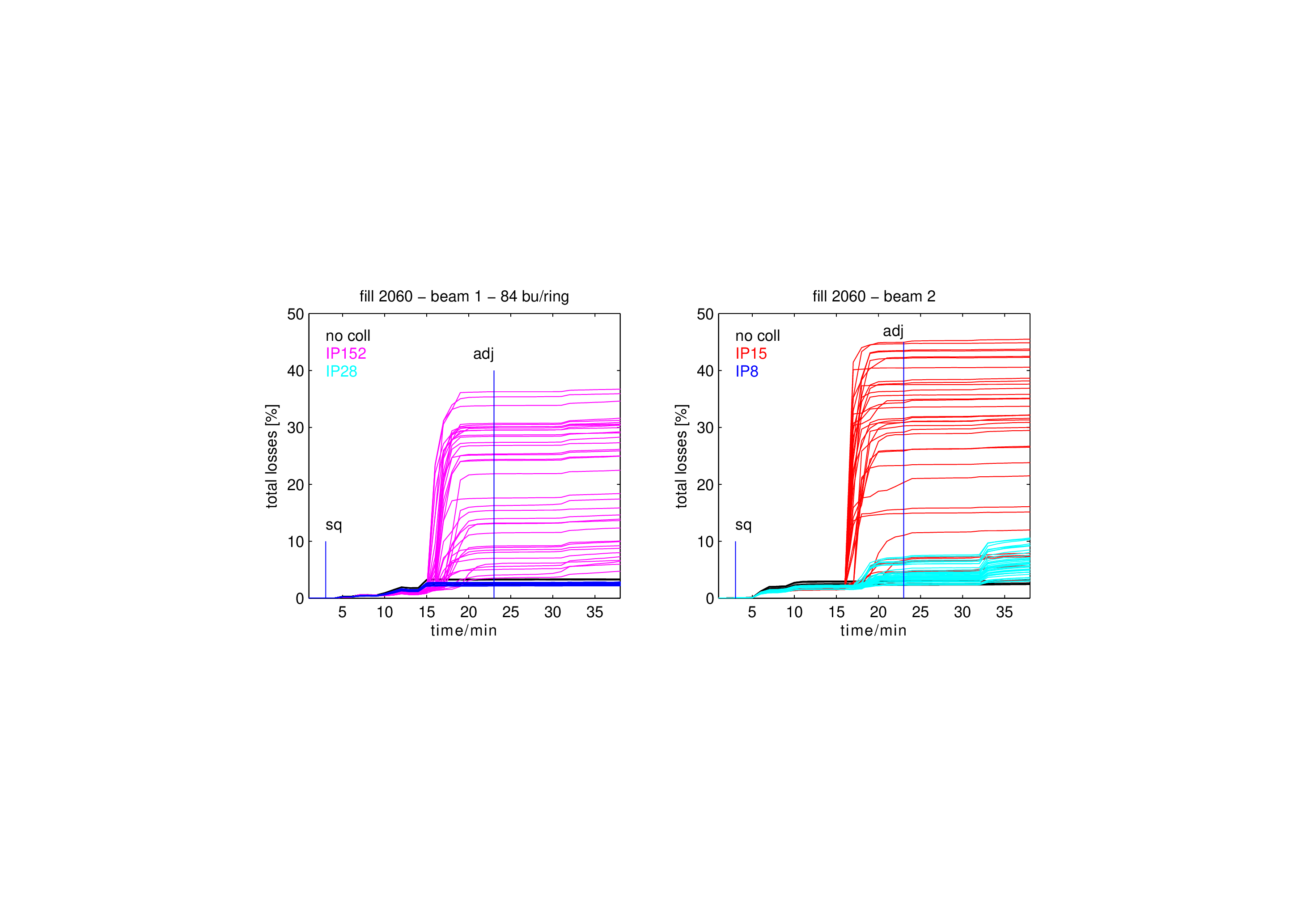}
   \caption{Losses in Beam 1 and Beam 2 with too small separation.}
   \label{ref-f3a}
\end{figure}

For the first attempt to squeeze the optical functions from $\beta^{*}~$=~1.5~m
to $\beta^{*}~$=~1.0~m, the crossing angle was decreased to reduce the
required aperture, thus reducing the separation at the encounters.
During the ramp, an instability occurred which (probably) increased the
emittances of all bunches, reducing further the normalized beam
separation.
When the optics was changed, very significant beam losses occurred (see Fig.~\ref{ref-f3a})
for those bunches colliding in interaction points 1 and 5, where the separation was reduced due to
smaller $\beta^{*}~$.
Bunches colliding only in interaction points 2 and 8 are not affected.

This clearly demonstrates the strong effect of long-range encounter and the
need for sufficient separation.
~~\
\subsection{Losses due to Long-range Encounters during Dedicated Experiments}
\begin{figure}[htb]
   \centering
   \includegraphics*[width=35mm, height=75mm, angle=-90]{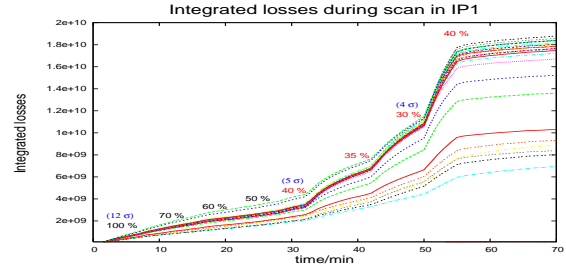}
   \caption{Integrated losses of all bunches as a function of time during scan of beam separation in IP1. Numbers show percentage of full crossing angle.}
   \label{ref-f2}
\end{figure}

We have performed two measurements, and the
results of the first experiment are shown in Fig.~\ref{ref-f2}, where the
integrated losses for the 36 bunches in Beam 1 are shown as a function of time
and the relative change of the crossing angle is given as a percentage of the
nominal (100\%~$\equiv$~240~$\mu$rad). The nominal value corresponds to a separation
of approximately 12~$\sigma$ at the parasitic encounters.

\begin{figure}[htb]
   \centering
\includegraphics[height= 3.0cm,width= 8.0cm,angle=-00]{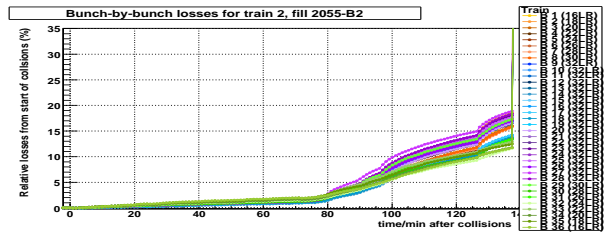}
   \caption{Integrated losses of bunches as a function of time during scan of beam separation in IP1. Bunches colliding in IP1 and IP5. Numbers show percentage of full crossing angle.}
   \label{ref-f2a}
\end{figure}

The losses per bunch observed in the second experiment are shown in Fig.~\ref{ref-f2a}.
The observed behaviour is very similar.

\begin{figure}[htb]
   \centering
\includegraphics[height= 3.0cm,width= 8.0cm,angle=-00]{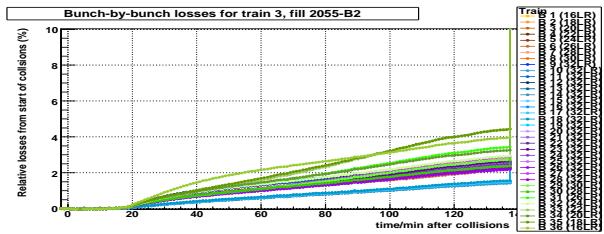}
   \caption{Integrated losses of bunches as a function of time during scan of beam separation in IP1. Bunches without collisions in IP1 and IP5. Numbers show percentage of full crossing angle.}
   \label{ref-f2aa}
\end{figure}

In this experiment we have set up several trains with different collision schemes and in
Fig.~\ref{ref-f2aa} we show the losses in the bunches colliding in IP8, but not in IP1 and IP5
where the separation was reduced.
As expected, this reduction had no effect on the losses of these bunches (please note change  of the scale).

\begin{figure}[htb]
   \centering
\includegraphics[height= 3.0cm,width= 8.0cm,angle=-00]{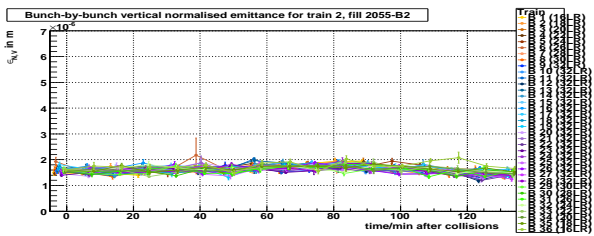}
   \caption{Vertical emittances all bunches as a function of time during scan of beam separation in IP1. }
   \label{ref-f2b}
\end{figure}

In Fig.~\ref{ref-f2b} we show the vertical emittances for all bunches during this second experiment.
Such a measurement was not available in the first study.

From Fig.~\ref{ref-f2} and Fig.~\ref{ref-f2a} we observe significantly increased losses
for some bunches when the separation is reduced to about 40\%, i.e.\ around 5~$\sigma$.
The emittances are not affected by the reduced separation (Fig.~\ref{ref-f2b}) and we interpret
this as a reduction of the dynamic aperture as expected from the theory and simulations.
The emittances, mainly determined by the core of the beam, are not affected in this case.
\subsection{Bunch to Bunch Differences and PACMAN Effects}
\begin{figure}[htb]
   \centering
   \includegraphics*[width=30mm, height=75mm, angle=-90]{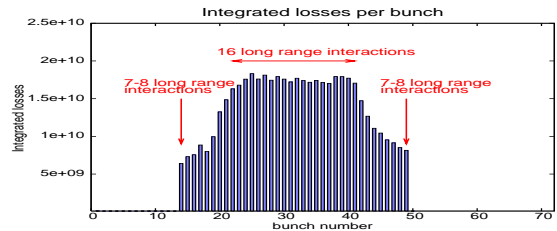}
   \caption{Integrated losses of all bunches along a train of 36 bunches, after reducing the crossing angle in IP1.}
   \label{ref-f1}
\end{figure}

Not all bunches are equally affected. At a smaller separation of 30\%, all bunches
experience significant losses ($\approx $4 $\sigma$).
Returning to a separation of 40\% reduces the losses significantly, suggesting
that mainly particles at large amplitudes have been lost during the scan due to
a reduced dynamic aperture.
Such behaviour is expected \cite{wh:05, wh:04}.
The different behaviour is interpreted as a "PACMAN" effect and should depend
on the number of long-range encounters, which varies along the train.
This is demonstrated in Fig.~\ref{ref-f1} where we show the integrated losses
for the 36 bunches in the train at the end of the experiment.

The maximum loss is clearly observed for the bunches in the centre
of the train with the maximum number
of long-range interactions (16) and the losses decrease as the number of
parasitic encounters decrease.
The smallest loss is found for bunches with the minimum number of interactions,
i.e.\ bunches at the beginning and end of the train \cite{wh:01, wh:02}.
\begin{figure}[htb]
   \centering
   \includegraphics*[width=70mm, height=35mm, angle=-00]{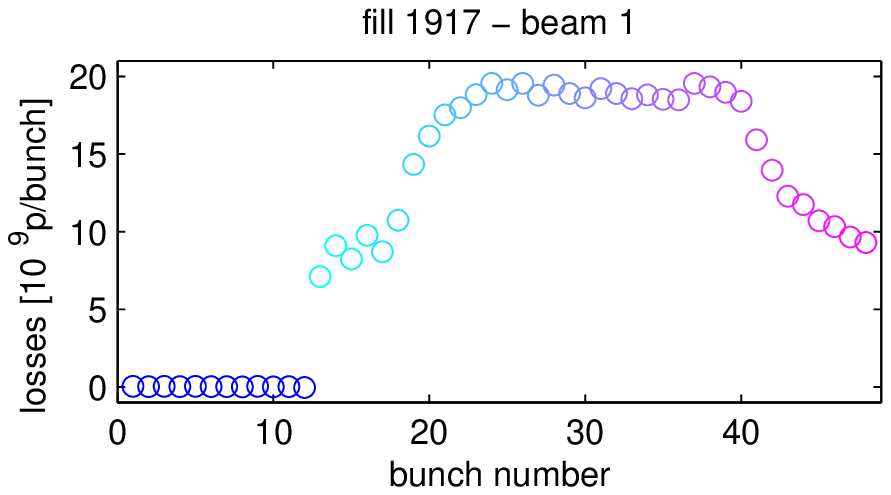}
   \includegraphics*[width=70mm, height=35mm, angle=-00]{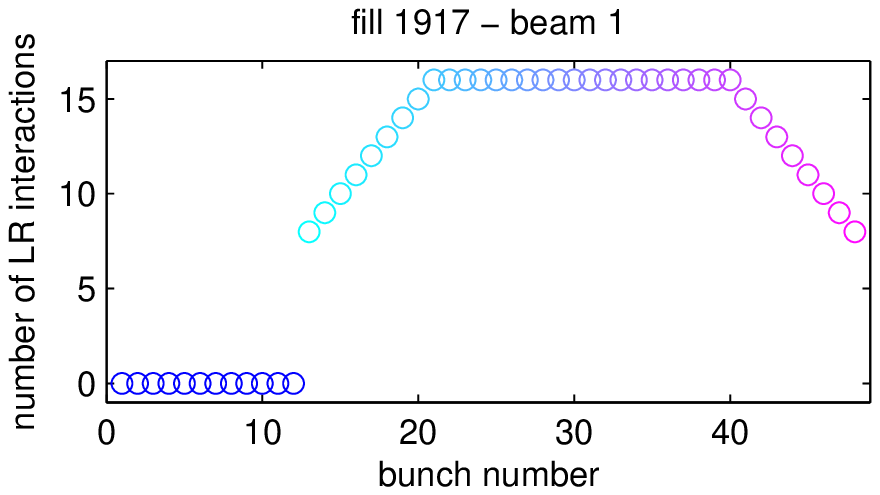}
   \caption{Integrated losses of all bunches along a train of 36 bunches, after reducing the crossing angle in IP1. The second figure shows number of long-range encounters for the same bunches.}
   \label{ref-f1c}
\end{figure}

This is demonstrated in Fig.~\ref{ref-f1c}, where we show the integrated losses, and in the
second figure for the same bunches the number of long-range encounters.
The agreement is rather obvious.
This is a very clear demonstration of the expected different behaviour, depending
on the number of interactions.

In the second part of the experiment we kept the separation at 40\% in IP1 and
started to reduce the crossing angle in the collision point IP5, opposite in
azimuth to IP1 (Fig.~\ref{ref-f0}).
Due to this geometry, the same pairs of bunches meet at the interaction
points, but the long-range separation is in the orthogonal plane.
This alternating crossing scheme was designed to compensate
first-order effects from long-range interactions \cite{wh:01}.

\begin{figure}[htb]
   \centering
   \includegraphics*[width=35mm, height=75mm, angle=-90]{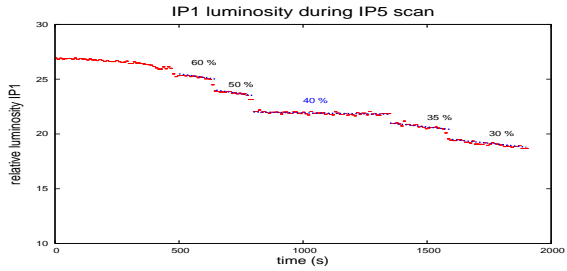}
   \caption{Luminosity in IP1 as a function of time during scan of beam separation in IP5.}
   \label{ref-f3}
\end{figure}

\begin{figure}[htb]
   \centering
   \includegraphics*[height= 8.2cm,width= 3.2cm,angle=-90]{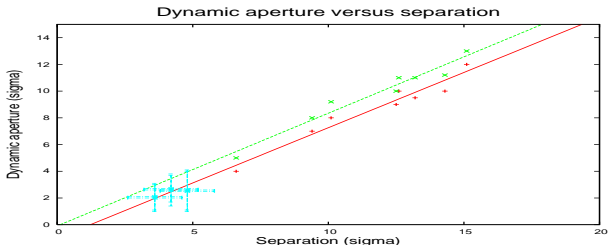}
   \caption{Dynamic aperture versus separation. Comparison with results from experiment.}
   \label{ref-f3b}
\end{figure}

Figure~\ref{ref-f3} shows the evolution of the luminosity in IP1 as we
performed the scan in IP5. The numbers indicate again the relative change of
separation, this time the horizontal crossing angle in IP5.
The luminosity seems to show that the lifetime is best when
the separation and crossing angles are equal for the two collision points.
It is worse for smaller as well as for larger separation.
This is the expected behaviour for a passive compensation due to alternating
crossing planes, although further studies are required to
conclude.
A more quantitative comparison with the expectations is shown in Fig.~\ref{ref-f3b}.
The dynamic aperture in units of the beam size is shown as a function of the
beam separation \cite{wh:05}.

From the relative losses in the experimental studies, we have tried to estimate the
dynamic aperture, assuming a Gaussian beam profile and tails.
This measurement can obviously only give a rough estimate, but is in very good
agreement with the expectations.
At larger separation, the losses are too small to get a reasonable estimate.
More information can be obtained from an analytical model \cite{wh:06, wh:07}.
\subsection{Further Observations of PACMAN Effects}
\begin{figure}[htb]
   \centering
   \includegraphics*[width=75mm, height=30mm, angle=-00]{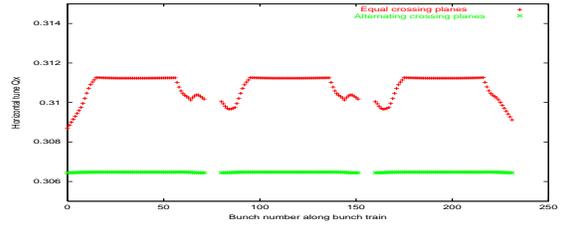}
   \caption{Computed tune along bunch train for equal and alternating crossing planes \cite{wh:01, wh:02}.}
   \label{ref-f4_0}
\end{figure}

The behaviour of so-called PACMAN bunches \cite{wh:02, dn:01} was always a
concern in the design of the LHC.
In order to avoid a tune shift of PACMAN bunches relative to the nominal bunches,
an alternating crossing scheme was implemented in the LHC \cite{wh:02}.
The effect of the alternating crossing scheme on the tune along a bunch train is shown in
Fig.~\ref{ref-f4_0}.
The computation is based on a self-consistent calculation of orbits and all optical
beam parameters \cite{wh:03}.

Without the alternating crossing, the PACMAN bunches exhibit a strong dependence of their
tunes on their position in the bunch train.
Depending on the intensity, bunch spacing and separation, this spread can exceed $2\times 10^{-3}$.
The alternating crossing scheme compensates completely for this spread.
This compensation is incomplete when bunch to bunch fluctuations are taken into account,
but in all cases the compensation is efficient \cite{wh:02}.
This compensation is largely helped by the design feature that the two low $\beta^{*}$
experimental regions are exactly opposite in azimuth (see Fig.~\ref{ref-f0}) \cite{gp:01}
and the same bunch pair collide in the two regions with alternating crossings and the
same optical parameters.
This requires that the contribution to the long-range beam--beam effects from the
other two experiments is small.
Due to the larger $\beta^{*}$ this is guaranteed under nominal operational conditions.

\begin{figure}[htb]
   \centering
   \includegraphics*[width=75mm, height=30mm, angle=-00]{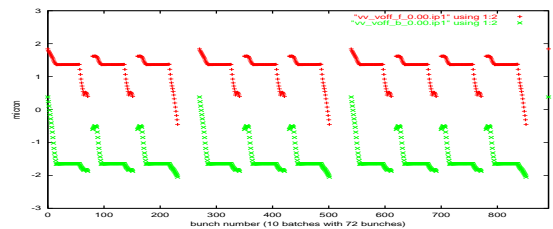}
   \caption{Computed orbit offsets in IP1 for Beam 1 and Beam 2 two vertical crossings \cite{wh:01, wh:02, wh:03}.}
   \label{ref-f4_1}
\end{figure}

\begin{figure}[htb]
   \centering
   \includegraphics*[width=75mm, height=30mm, angle=-00]{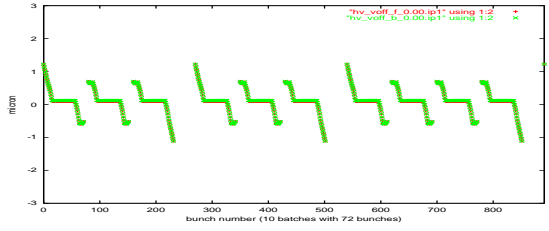}
   \caption{Computed orbit offsets in IP1 for Beam 1 and Beam 2 alternating crossings \cite{wh:01, wh:02, wh:03}.}
   \label{ref-f4_2}
\end{figure}

Another predicted behaviour of PACMAN bunches are the different orbits due
to the long-range interactions.
The decreased separation, corresponding to stronger dipolar kicks, clearly lead
to orbit changes along the corresponding bunch train.
The bunches not participating in collisions in IP1 and IP5 are not affected.
To study these effects, a fully self-consistent treatment was developed to compute
the orbits and tunes for all bunches in the machine under the influence of the
strong long-range beam--beam interactions \cite{wh:03}.
Figures~\ref{ref-f4_1} and \ref{ref-f4_2} show the vertical orbit offsets at IP1
for the two beams in the case of vertical crossings in IP1 and IP5.

The figures show a significant effect at the interaction point, and, moving the beams,
it is not possible to make all bunches overlap.

The effect of alternating crossings (vertical in IP1 and horizontal in IP5) is shown in
Fig.~\ref{ref-f4_2}.
Now the bunches from the two beams can be made overlap exactly, although not at the central collision
point.
In the other plane this complete overlap cannot be obtained, although the offset is small.

\begin{figure}[htb]
   \centering
   \includegraphics*[width=30mm, height=75mm, angle=-90]{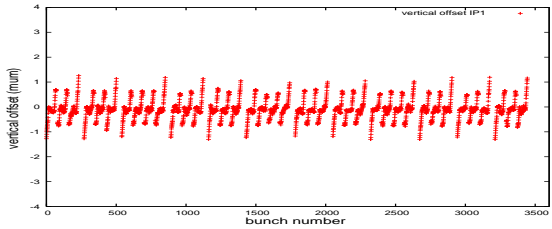}
   \caption{Computed orbit offsets in IP1 along the bunch train \cite{wh:01, wh:02}.}
   \label{ref-f4}
\end{figure}

In Fig.~\ref{ref-f4} we show a prediction for the vertical offsets in IP1 \cite{wh:01, wh:02}.
The offsets should vary along the bunch train.
Although the orbit measurement in the LHC is not able to resolve these effects,
the vertex centroid can be measured bunch by bunch in the experiment.

\begin{figure}[htb]
   \centering
   \includegraphics*[width=75mm, height=40mm, angle=-0]{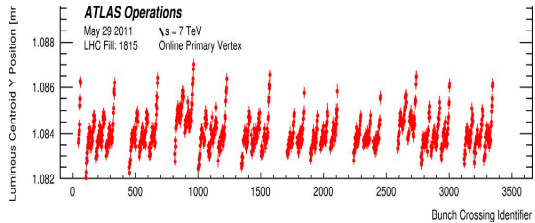}
   \caption{Measured orbit offsets in IP1 along the bunch train \cite{wk:01, wk:02}.}
   \label{ref-f5}
\end{figure}

The measured orbit in IP1 (ATLAS experiment) is shown in Fig.~\ref{ref-f5}
and at least the qualitative agreement is excellent.
This is a further strong indication that the expected PACMAN effects are present and
understood and that our computations are reliable.
\section{PARAMETRIC DEPENDENCE OF LONG-RANGE LOSSES}
In order to study the dependence of long-range effects on the parameters of the
beam--beam interaction, we have performed the experiments with different parameters,
in particular different $\beta^{*}$ and intensities.
The relevant parameters of the three experiments are found in Tab.~\ref{tab:01}.

\begin{table} [htb]
\caption{Parameters for three long-range experiments}
   \label{tab:01}
\centering
\begin{tabular}{llll}\hline\hline
        \textbf{Experiment}           & \textbf{Emittance}  &  $\beta^{*}$   &  \textbf{Intensity}  \\   \hline
2011 (50 ns) &  2.0 - 2.5 $\mu$m  &  1.5 m &1.2 10$^{11}$   \\  \hline
2012 (50 ns) &  2.0 - 2.5 $\mu$m  &  0.6 m &1.2 10$^{11}$   \\  \hline
2012 (50 ns) &  2.0 - 2.5 $\mu$m  &  0.6 m &1.6 10$^{11}$   \\  \hline\hline
\end{tabular}
\end{table}

The experimental procedure was the same as before: the separation (crossing angle) was reduced
until visible losses were observed.

\begin{figure}[htb]
{\centering{\includegraphics[height= 4.0cm,width=90.0mm,angle=-00]{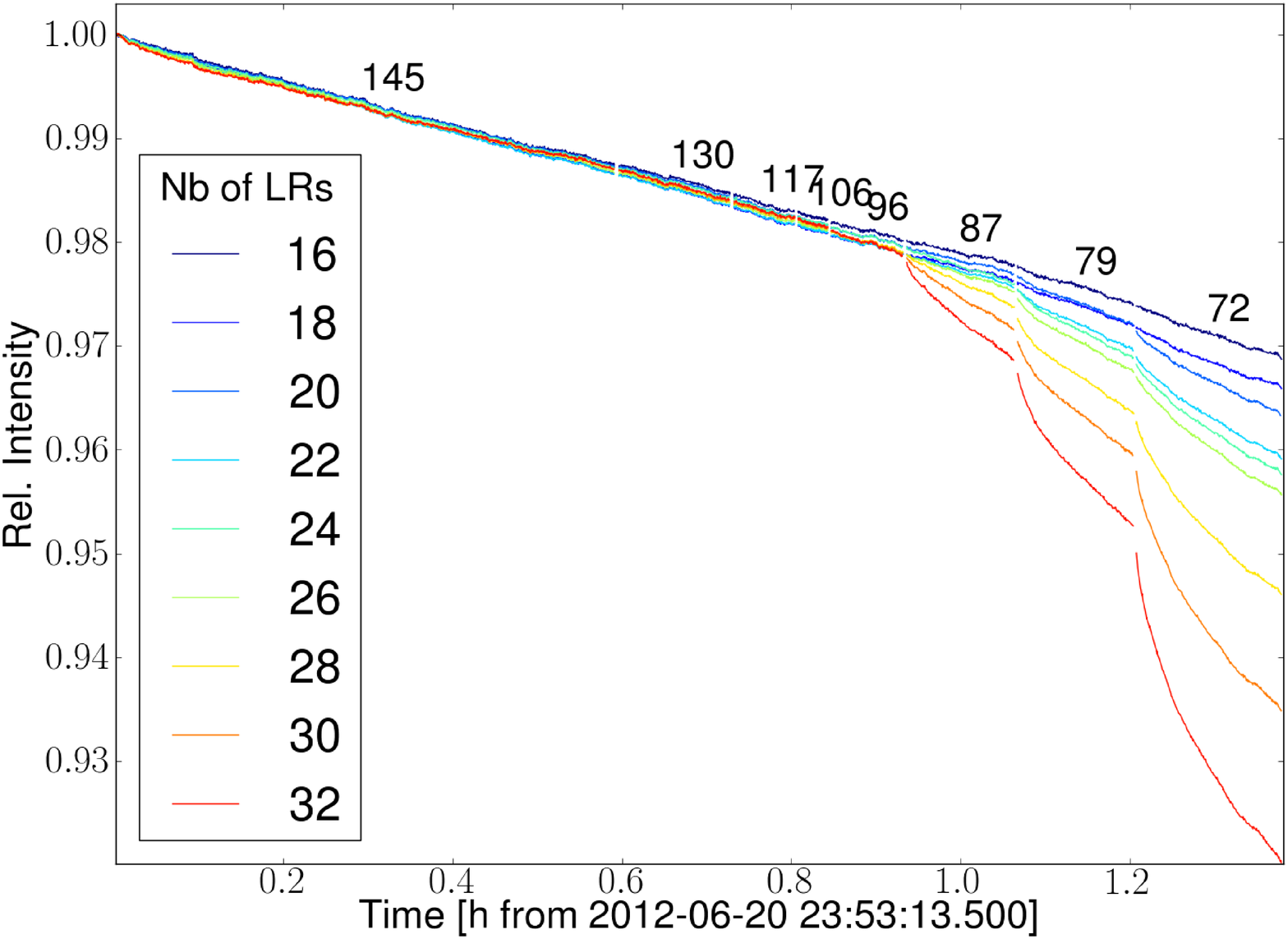}}}
   \caption{Separation scan with high intensity.}
   \label{ref-s1}
\end{figure}
The results of a first separation scan are shown in Fig.~\ref{ref-s1}.
The main observations are as follows:
\begin{itemize}
\item[$\bullet$] Recent test (2012) with {{$\beta^{*}$ = 0.60m}}, intensity: {{1.6~10$^{11}$~p/bunch}}
\item[$\bullet$] Initial beam separation $\approx$ 9 - 9.5 $\sigma$
\item[$\bullet$] Losses start at $\approx$ 6 $\sigma$ separation
\end{itemize}

\begin{figure}[htb]
{\centering{\includegraphics[height= 4.0cm,width=90.0mm,angle=-00]{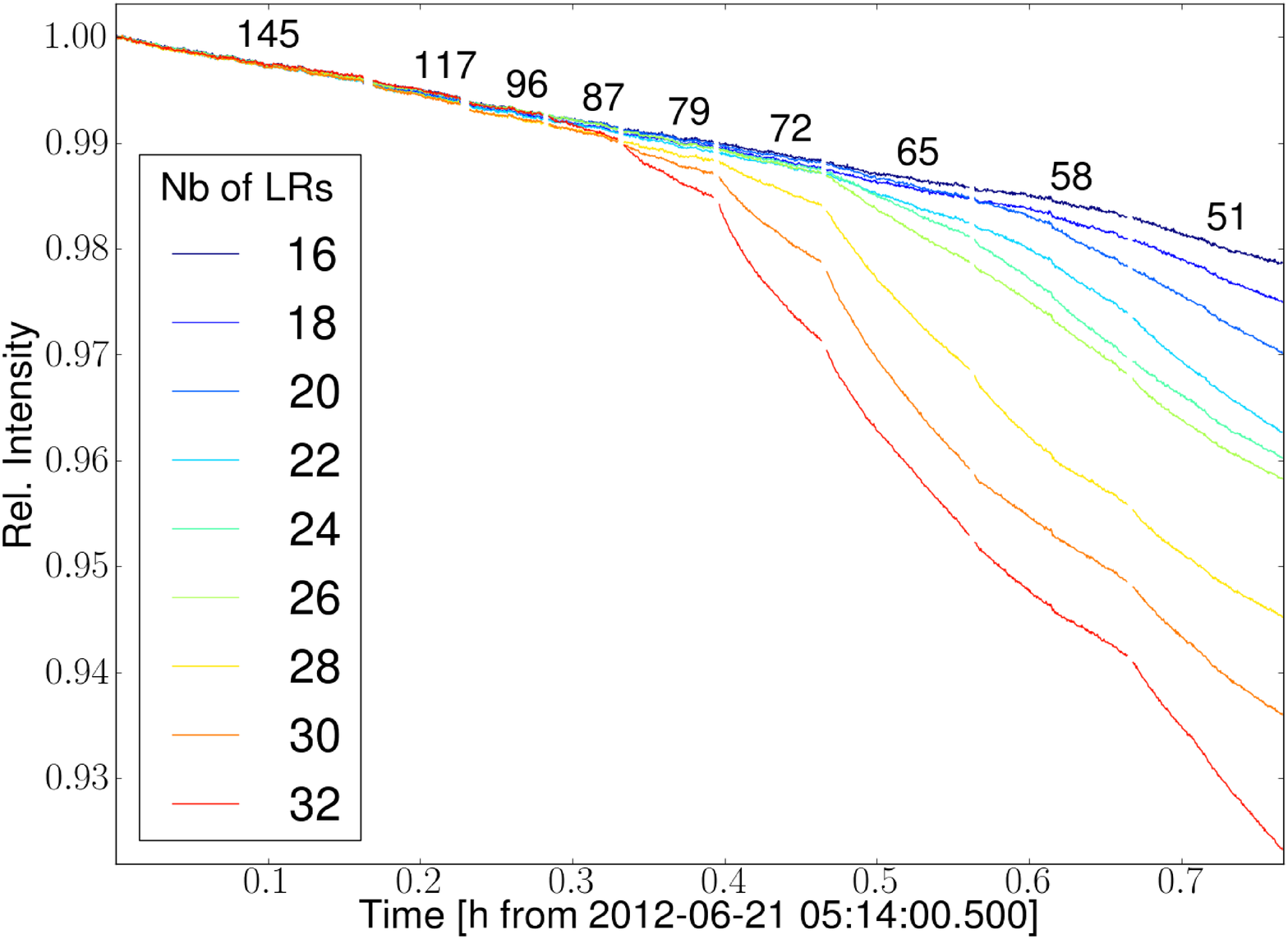}}}
   \caption{Separation scan with reduced intensity.}
   \label{ref-s2}
\end{figure}
The results of a second separation scan are shown in Fig.~\ref{ref-s2}.
The main observations are as follows:
\begin{itemize}
\item[$\bullet$] Recent test (2012) with {{$\beta^{*}$ = 0.60~m}}, intensity: {{1.2~10$^{11}$~p/bunch}}.
\item[$\bullet$] Initial beam separation $\approx$ 9 - 9.5 $\sigma$.
\item[$\bullet$] Losses start at $\approx$ 5 $\sigma$ separation.
\end{itemize}
The experiments summarized in Tab.~\ref{tab:01} have been analysed
using a recently developed technique to parametrize the strength
of the long-range non-linearity, based on the evaluation of the
invariant and the smear \cite{wh:06, wh:07}.

This method is applied to compare different configurations \cite{wh:07}
and allows us to derive scaling laws for the dynamic aperture.
\section{SUMMARY}
We have reported on the first studies of beam--beam effects in the LHC with
high intensity, high brightness beams and can summarize the results as
follows.
\begin{itemize}
\item[$\bullet$] The effect of the beam--beam interaction on the beam dynamics is clearly established.
\item[$\bullet$] The effect of long-range interactions on the beam lifetime and losses (dynamic aperture) is clearly visible.
\item[$\bullet$] The number of head-on and/or long-range interactions important for losses and all predicted PACMAN effects are observed.
\end{itemize}
All observations are in good agreement with the expectations and an analytical
model \cite{wh:07}.
From this first experience we have confidence that beam--beam effects in the LHC are
understood and should allow us to reach the target luminosity for the nominal
machine at 7~TeV beam energy.
The analytical model \cite{wh:07} should allow us to extrapolate the results to
different configurations and allow an optimization of the relevant parameters.
\section{ACKNOWLEDGEMENTS}
These studies would have been impossible without the help and support from
the operations teams for the LHC and its injectors.
We are also grateful to the LHC experiments for the collaboration during
our studies and for providing us with luminosity and background data.

\end{document}